\documentstyle[twoside,fleqn,espcrc2,epsf]{article}
\input psfig
\def\gsim{\mathrel{\raise.3ex\hbox{$>$}\mkern-14mu
             \lower0.6ex\hbox{$\sim$}}}
\def\lsim{\mathrel{\raise.3ex\hbox{$<$}\mkern-14mu
             \lower0.6ex\hbox{$\sim$}}}

\newcommand{\AmS}{{\protect\the\textfont2
  A\kern-.1667em\lower.5ex\hbox{M}\kern-.125emS}}

\title{Radical Conservatism And Nucleon Decay}

\author{Frank Wilczek\address{Institute for Advanced Study,\\ School
        of Natural Sciences, Olden Lane, \\ Princeton, New Jersey
        08540} \thanks{Invited Talk at NNN99 Workshop, September
        23-25, 1999, at SUNY-Stony Brook.  IASSNS-HEP 00/09. Research
        supported in part by DOE grant
        DE-FG02-90ER40542. }} 

\begin{document}


\begin{abstract}
Unification of couplings, observation of neutrino masses in the
expected range, and several other considerations confirm central
implications of straightforward gauge unification based on $SO(10)$ or
a close relative and incorporating low-energy supersymmetry.  The
remaining outstanding consequence of this circle of ideas, yet to be
observed, is nucleon instability.  Clearly, we should aspire to be as
specific as possible regarding the rate and form of such instability.
I argue that not only esthetics, but also the observed precision of
unification of couplings, favors an economical symmetry-breaking
(Higgs) structure.  Assuming this, one can exploit its constraints to
build reasonably economical, overconstrained yet phenomenologically
viable models of quark and lepton masses.  Putting it all together,
one arrives at reasonably concrete, hopeful expectations regarding
nucleon decay.  These expectations are neither ruled out by existing
experiments, nor hopelessly inaccessible.
\end{abstract}
\maketitle
Radical conservatism, in the sense of Wheeler, is the doctrine of
taking good successful ideas seriously, and pressing them hard, to see
if they break.  It has a noble history, for example in quantum
electrodynamics .  Here I will follow this philosophy for
straightforward gauge unification.  In the recent literature many more
exotic ideas about physics beyond the standard model have been
explored \cite{bachas}, and there is nothing wrong with that, but one
should not forget that the simplest possibilities, already broadly
envisioned by the early 80s, have not been disproved.  Quite the
contrary.  For reasons I will presently summarize, I believe that
after years of marvelous precision work at LEP and elsewhere, the
discovery of non-zero neutrino mass at SuperK, and the non-discovery
of any among a plethora of suggested exotica, the early ideas look
better than ever.  Maybe it is a coincidence -- excuse me, a series of
coincidences -- or a conspiracy.  Maybe.  But I doubt it, and so
should you.

This is not to say that gauge theory unification is the end of all
desire, or a Theory of Everything.  It certainly is not.  Even if
true, it leaves many loose ends and unanswered questions.  But if true
it represents a worthy addition to the Standard Model, a major
additional insight into Nature, and a foundation for further progress.

And, most fortunately, gauge theory unification is quite concretely a
theory of Something.  In many ways the crown jewel among its
predictions is that nucleons should decay.  The possibility of such
decay directly reflects the unity of matter -- interconvertibility of
quarks and leptons -- and connects to the cosmological asymmetry
between matter and antimatter.  The quest to observe nucleon decay has
already inspired heroic, though so far fruitless, experimental
efforts.

Actually, after a moment's reflection, I want to take back that `so far
fruitless'.  Creative efforts to observe nucleon decay have led, through
the great IMB, Kamiokande, and SuperK lineage of experiments, to
technology that has proved immensely fruitful for neutrino
physics. Highlights include observation of the supernova 1987a burst,
observation of oscillations in neutrinos deriving from atmospheric
cosmic rays \cite{superK}, and observation of a non-zero but anomalous 
high-energy
solar neutrino flux -- each of these representing an achievement of
historic proportions.  And even the negative result of nucleon
instability searches 
to date has been of genuine positive value.  It provided an
early motivation for supersymmetric unification, and continues to
offer powerful guidance as to what proposals for physics beyond the
standard model can be considered plausible.

In any case, we are gathered here to consider whether still more
heroic, not to mention expensive, efforts in this direction are
warranted.  And I want to argue as forcefully as I can for what I
believe, that they most certainly are.  For upon putting together a
number of elegant, successful ideas one arrives at reasonably
concrete, hopeful expectations regarding nucleon decay, as I shall
indicate.  These expectations are neither hopelessly inaccessible, nor
ruled out by existing experiments.  Furthermore, the branching
fractions can discriminate among different possibilities for physics
at the unification scale.  I will be drawing on results from a recent
long, numerically dense analysis by K. Babu, J. Pati, and myself \cite{bpw}.
This work in turn draws on an extensive previous literature; for details
and references you should refer to our paper.

\section{The Case for Unification} 
The argument for gauge unification is powerful and many-faceted.  I
will review it in seven installments, starting with the strongest and
working down: 
\begin{enumerate}
\item the unification of quantum numbers and multiplets;

\item the unification of couplings, using supersymmetry; 

\item the
explanation of small neutrino masses, in the observed range; 

\item the explanation of the b/$\tau$ mass ratio; 

\item explaining why things that might otherwise happen, do not; 

\item propinquity of the unification and quantum gravitational scales; 

\item broad consistency with string/M theory.
\end{enumerate}

\subsection{quantum numbers and multiplets}

\begin{figure}[htb]
\centerline{\psfig{figure=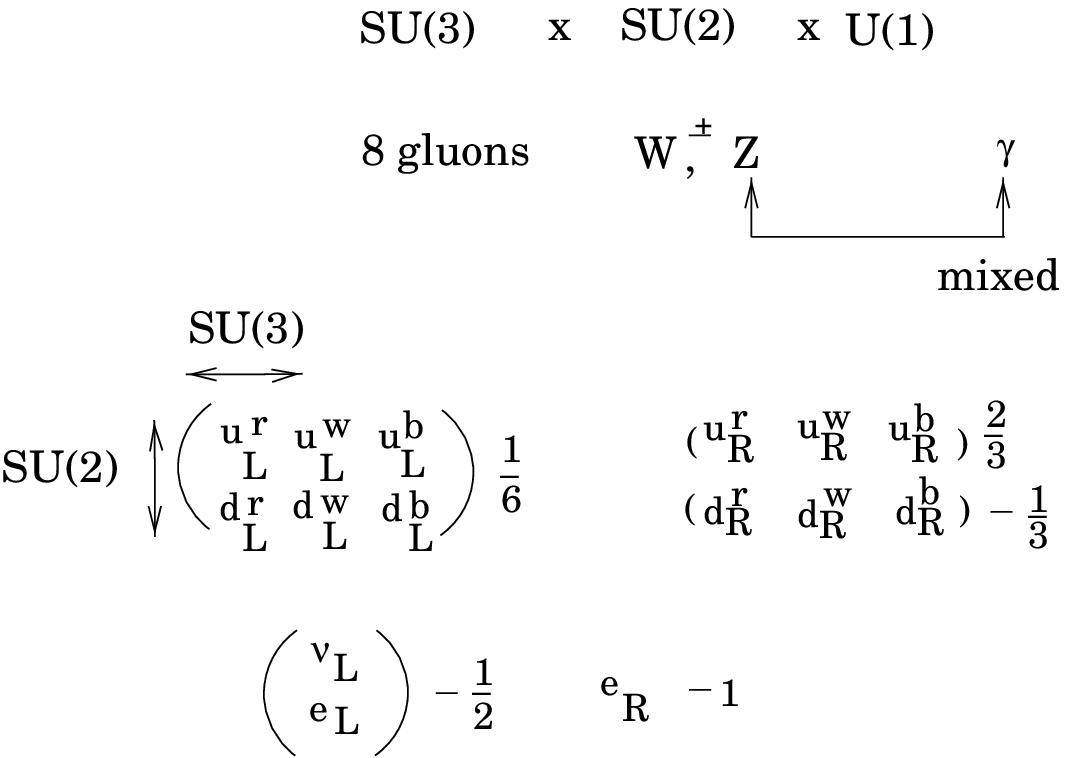,width=7cm}}
\vglue-.4in
\caption[]{The gauge groups of the standard model, and the
fermion multiplets with their hypercharges.}
\label{fig:su1.eps}
\end{figure}

The standard model of particle physics is based upon the
gauge groups \break\hfil 
SU(3)$\times$SU(2)$\times$U(1) of strong, electromagnetic
and weak interactions acting on the quark and lepton multiplets
as shown in Figure \ref{fig:su1.eps}.

In this Figure I have depicted only one family (u,d,e,$\nu_e$) of
quarks and leptons; in reality there seem to be three families that
are mere copies of one another as far as their interactions with the
gauge bosons are concerned, but differ in mass.  Actually in the
Figure I have ignored masses altogether, and allowed myself the
convenient fiction of pretending that the quarks and leptons have a
definite chirality -- right- or left-handed -- as they would if they
were massless.  The more precise statement, of course, is that the
gauge bosons couple to currents of definite chirality.  The chirality
is indicated by a subscript R or L.  Finally the little number beside
each multiplet is its assignment under the U(1) of hypercharge, which
is the average of the electric charge of the multiplet.

While little doubt can remain that the Standard Model is essentially
correct, a glance at Figure \ref{fig:su1.eps} is enough to reveal that
it is not a complete or final theory.  To remove its imperfections,
while building upon its solid success, is a worthy challenge.

Given that the strong interactions are governed by transformations
among three colors, and the weak by transformations between two
others, what could be more natural than to embed both theories into a
larger theory of transformations among all five colors?

This idea has the additional attraction that an extra U(1) symmetry
commuting with the strong SU(3) and weak SU(2) symmetries
automatically appears, which we can attempt to identify with the
remaining gauge symmetry of the standard model, that is hypercharge.
For while in the separate SU(3) and SU(2) theories we must throw out
the two gauge bosons which couple respectively to the color
combinations R+W+B and G+P, in the SU(5) theory we only project out
R+W+B+G+P, while the orthogonal, traceless combination
(R+W+B)-${3\over 2}$(G+P) remains.

Finally, the possibility of unified gauge symmetry breaking is
plausible by analogy; after all, we know for sure that gauge symmetry
breaking occurs in the electroweak sector.

Georgi and Glashow \cite{gg}, (and in a different way, Pati and Salam
\cite{unif}) showed how these ideas can be used to bring some order to
the quark and lepton sector, and in particular to supply a satisfying
explanation of the weird hypercharge assignments in the standard
model.  As shown in Figure \ref{fig:colors}, the five scattered
SU(3)$\times$SU(2)$\times$U(1) multiplets get organized into just two
representations of SU(5).

In making this unification it is necessary to allow transformations
between (what were previously considered to be) particles and
antiparticles, and also between quarks and leptons.  It is convenient
to work with left-handed fields only.  Since the conjugate of a
right-handed field is left-handed, we don't lose anything by doing so
-- though we must shed traditional prejudices about a rigorous
distinction between matter and antimatter, since these get mixed up.
Specifically, it will not be possible to declare that matter is what
carries positive baryon and lepton number, since the unified theory
does not conserve these quantum numbers.

As shown in Figure \ref{fig:colors}, there is one group of ten
left-handed fermions that have all possible combinations of one unit
of each of two different colors, and another group of five left-handed
fermions that each carry just one negative unit of some color.  These
are the ten-dimensional antisymmetric tensor and the complex conjugate
of the five-dimensional vector representation, commonly referred to as
the five-bar.  In this way, {\it the structure of the standard model,
with the particle assignments gleaned from decades of experimental
effort and theoretical interpretation, is perfectly reproduced by a
simple abstract set of rules for manipulating symmetrical symbols}.
Thus for example the object RB in this Figure has just the strong,
electromagnetic, and weak interactions we expect of the complex
conjugate of the right-handed up-quark, without our having to instruct
the theory further.

A most impressive, though simple, exercise is to work out the
hypercharges of the objects in Figure \ref{fig:colors} and checking
against what you need in the Standard Model.  These ugly ducklings of
the Standard Model have matured into quite lovely swans.

\bigskip

\begin{figure}
\parskip=0pt{
\underline{SU(5):  5 colors RWBGP}

$\underline{10}$: 2 different color labels (antisymmetric tensor)

$$\matrix{\rm u_L:&\rm RP,&\rm WP,&\rm BP\cr
\rm d_L:&\rm RG,&\rm WG,&\rm BG\cr
\rm u{^c_L}:&\rm RW,&\rm WB,&\rm BR\cr
&\rm (\bar B)&\rm (\bar R)&\rm (\bar W)\cr
\rm e{^c_L}:&\rm GP&&\cr
&(\ )&&\cr
}
\pmatrix{0&\rm u^c&\rm u^c&\rm u&\rm d\cr
&0&\rm u^c&\rm u&\rm d\cr
&&0&\rm u&\rm d\cr
&*&&0&\rm e\cr
&&&&0\cr}$$

$\underline{\bar 5}$: 1 anticolor label

$$\matrix{\rm d{^c_L}:&\rm \bar R,&\rm  \bar W,&\rm  \bar B\cr
\rm e_L:&\rm \bar P&&\cr
\nu_{\rm L}:&\rm \bar G&&\cr
}
\matrix{\rm (d^c&\rm d^c&\rm d^c&{\rm e}&\nu)\cr}$$
\def\boxtext#1{%
\vbox{%
\hrule
\hbox{\strut \vrule{} #1 \vrule}%
\hrule
}%
}
\centerline{
\vbox{\offinterlineskip
\hbox{\boxtext{\rm Y $= -{1\over 3}$ (R+W+B) $+{1\over 2}$ (G+P)}}
}}
}
\vglue-.2in
\caption[]{Unification of fermions in SU(5)  There is a beautiful
extension of SU(5) to the slightly larger group SO(10).  With this
extension, one can unite all the observed fermions of a family, plus
one more, into a {\it single\/} multiplet \cite{so10}.  The relevant
representation for the fermions is a 16-dimensional spinor
representation.  Some of its features are depicted in Figure
\ref{fig:5bit}.
\label{fig:colors}}
\end{figure}

\begin{figure}
\parskip=0pt
\hglue0.75in\underline{SO(10): 5 bit register}
\vglue-.15in
$$(\pm \pm \pm \pm \pm)\ \ :\ \ \underline{\rm even}\ \  \# \  of\  -$$
$$10:\matrix{(++-|+-)&6&\rm (u_L,d_L)\cr
(+--|++)&3&\rm u{^c_L}\cr
(+++|--)&1&\rm e{^c_L}\cr}$$

$$\bar 5:\matrix{(+--|--)&\bar 3&\rm d{^c_L}\cr
(---|+-)&\bar 2&{\rm (e_L},\nu_L)\cr}$$
\nopagebreak
$$1:\matrix{(+++|++)&1&\rm N_R\cr}$$
\vglue-.2in
\caption[]{Unification of fermions in SO(10).  The rule is that all
possible combinations of 5 + and - signs occur, subject to the
constraint that the total number of - signs is even.  The SU(5) gauge
bosons within SO(10) do not change the numbers of signs, and one sees
the SU(5) multiplets emerging.  However there are additional
transformations in SO(10) but not in SU(5), which allow any fermion to
be transformed into any other.\label{fig:5bit}}
\end{figure}
 
In addition to the conventional quarks and leptons the SO(10) spinor
contains an additional particle, an SU(3)$\times$SU(2)$\times$U(1)
singlet.  (It is even an SU(5) singlet.)  Usually when a theory
predicts unobserved new particles they are an embarrassment.  But
these N particles -- there are three of them, one for each family --
are a notable exception.  Indeed, they are central to the emerging
connection between neutrino masses and unification, as I
shall discuss below.

\subsection{unification of couplings using supersymmetry}

We have just seen that simple unification schemes are spectacularly
successful at the level of classification.  New questions arise when
we consider dynamics.

Part of the power of gauge symmetry is that it fully dictates the
interactions of the gauge bosons, once an overall coupling constant
is specified.  Thus if SU(5) or some higher symmetry were exact, then
the fundamental
strengths of the different color-changing interactions would have
to be equal, as would the
(properly normalized) hypercharge coupling strength.  In reality the
coupling strengths of the gauge bosons in SU(3)$\times$SU(2)$\times$U(1)
are not observed to be equal, but rather follow the pattern
$g_3 \gg g_2 > g_1$.

Fortunately, experience with QCD emphasizes that couplings {\it 
run\/} \cite{dgfw}.
The physical mechanism of this effect is that in quantum field theory
the vacuum must be regarded as a polarizable medium, since virtual
particle-antiparticle pairs can screen charge.  For charged gauge
bosons, as arise in non-abelian theories, the paramagnetic
(antiscreening) effect of their spin-spin interaction dominates, which
leads to asymptotic freedom.  As Georgi, Quinn, and Weinberg pointed
out \cite{gqw}, if a gauge symmetry such as SU(5) is spontaneously broken
at some very short distance then we should not expect that the
effective couplings probed at much larger distances, such as are
actually measured at practical accelerators, will be equal.  Rather
they will all have been affected to a greater or lesser extent by
vacuum screening and anti-screening, starting from a common value at
the unification scale but then diverging from one another.  The
pattern $g_3 \gg g_2 > g_1$ is just what one should expect, since the
antiscreening effect of gauge bosons is more pronounced for larger
gauge groups.

\begin{figure}[!t]
\centerline{\psfig{figure=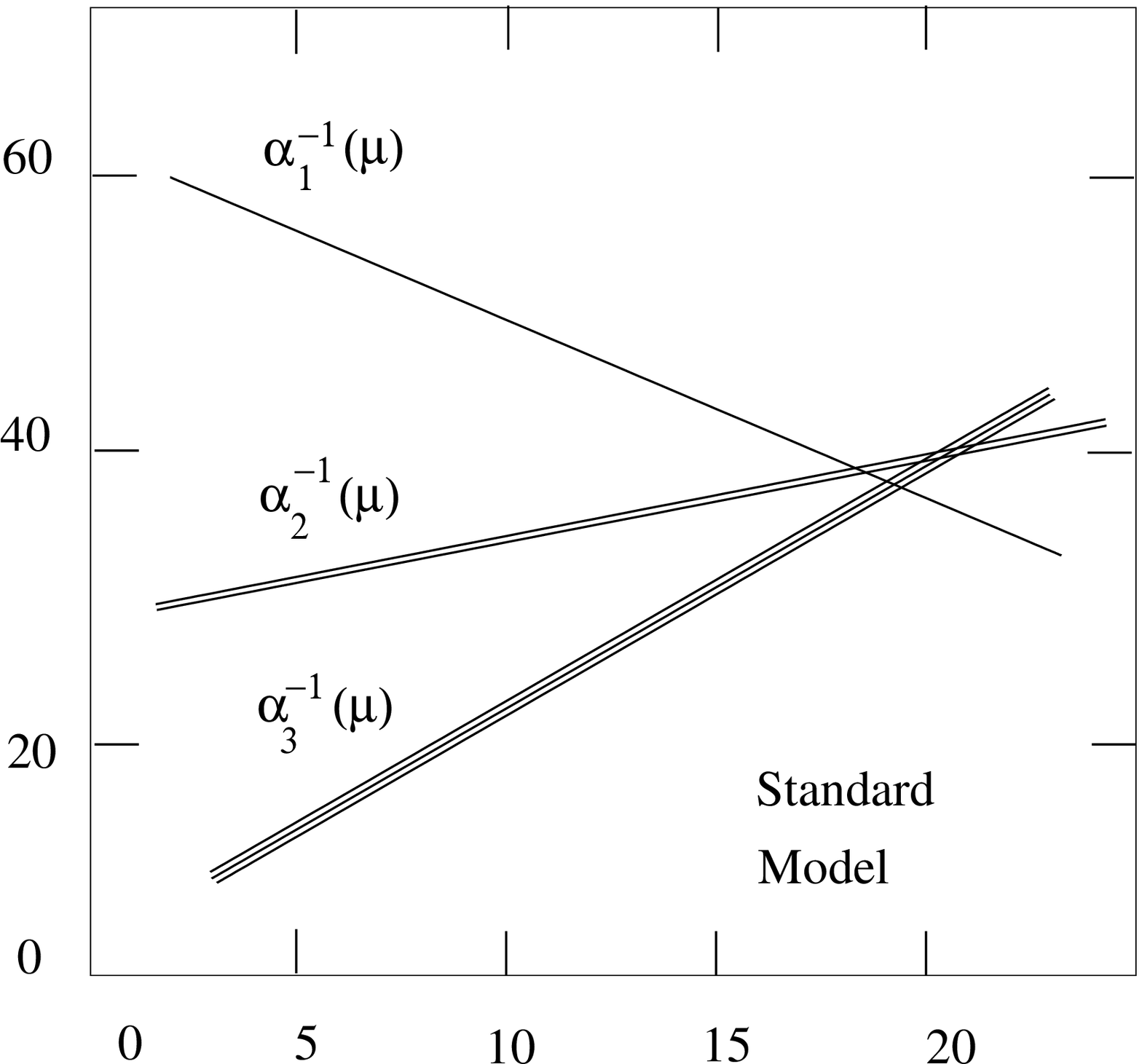,width=7cm}}
\vglue-.4in
\caption[]{The failure of the running couplings, normalized according
to SU(5) and extrapolated taking into account only the virtual
exchange of the ``known'' particles of the standard model (including
the  Higgs boson) to meet.  Note that only with fairly
recent experiments \cite{EWG}, which greatly improved the precision
of the determination of low-energy couplings, has the discrepancy
become significant.}
\label{fig:SM-run.eps}
\end{figure}

The running of the couplings gives us a truly quantitative handle on
the ideas of unification.  To specify the relevant aspects of
unification, one basically needs only to fix two parameters: the scale
at which the couplings unite, (which is essentially the scale at which
the unified symmetry breaks), and their common value when they unite.
Given these, one calculates three outputs, the three {\it a priori\/}
independent couplings for the gauge groups in
SU(3)$\times$SU(2)$\times$U(1).  Thus the framework is eminently
falsifiable.  The astonishing thing is, how close it comes to working
(See Figure 
\ref{fig:SM-run.eps}).

The GQW calculation is remarkably successful in
explaining the observed hierarchy $g_3 \gg g_2 > g_1$ of
couplings and the approximate stability of the proton.
In performing it, we assumed that the known and
confidently expected particles of the standard model exhaust
the spectrum up to the unification scale, and that the
rules of quantum field
theory could be extrapolated without alteration
up to this mass scale -- thirteen orders
of magnitude beyond the domain they were designed to describe.
It is a triumph for minimalism, both existential and conceptual.

On closer inspection, however, 
it is not quite good enough.  Accurate modern measurements
of the couplings show a small but definite discrepancy between the
couplings, as appears in Figure .
\ref{fig:SM-run.eps}.  
And heroic dedicated experiments to search for proton decay at the
rate expected from exchange of the additional gauge bosons present in
$SU(5)$ but not in the Standard Model did not find it \cite{blewitt}.
They currently exclude the minimal SU(5) prediction $\tau_p \sim
10^{31}~{\rm yrs.}$ by about two orders of magnitude.

If we just add particles in some haphazard
way things will
only get
worse: minimal SU(5) nearly works, so a generic perturbation
will be deleterious.  Even if some {\it ad hoc\/}
prescription could be made to work,
that would be a disappointing outcome from what
appeared to be one of our most precious, elegantly
straightforward clues regarding physics well
beyond the Standard Model.

\begin{figure}[!t]
\centerline{\psfig{figure=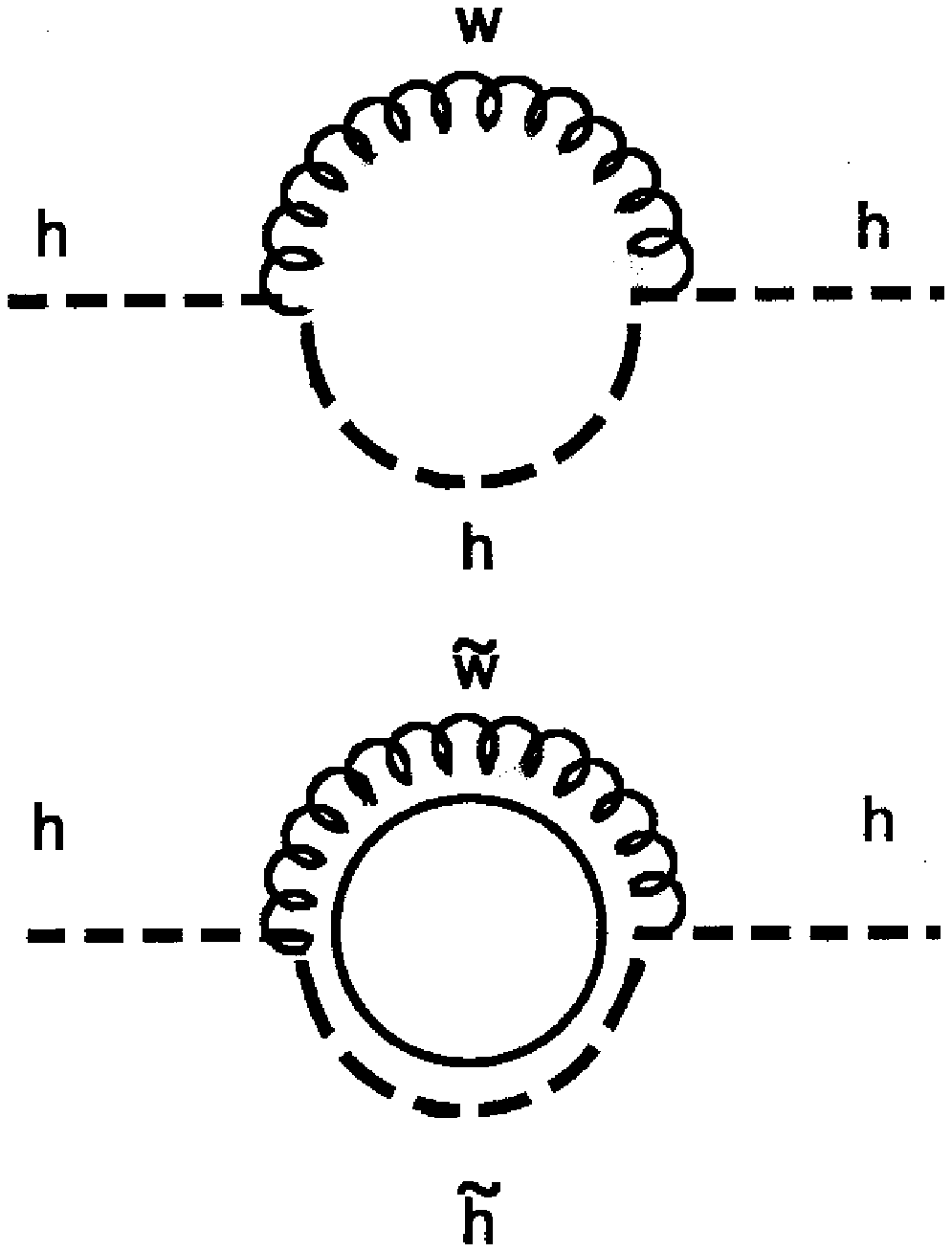,width=5cm}}
\caption[]{Contributions to the Higgs field self-energy.  These graphs
give contributions to the Higgs field self-energy which separately are
formally quadratically divergent, but when both are included the
divergence is removed.  In models with broken supersymmetry a finite
residual piece remains.  If one is to obtain an adequately small
finite contribution to the self-energy, the mass difference between
Standard Model particles and their superpartners cannot be too great.
This -- and essentially only this -- motivates the inclusion of
virtual superpartner contributions in Figure 6 
beginning at relatively low scales.}
\label{fig:cosfig8.eps}
\end{figure}

Fortunately, there is a compelling escape from this impasse.  That is
the idea of supersymmetry \cite{ferra}.  Supersymmetry is certainly
not a symmetry in nature: for example, there is certainly no bosonic
particle with the mass and charge of the electron.  However there are
several reasons for thinking that supersymmetry might be
spontaneously, and only relatively mildly broken, so that the
superpartners are no more massive than $\approx$ 1 Tev.  The most
concrete arises in calculating radiative corrections to the (mass)$^2$
of the Higgs particle from diagrams of the type shown in {}Figure
\ref{fig:cosfig8.eps}.  One finds that they make an infinite, and also
large, contribution.  By this I mean that the divergence is quadratic
in the ultraviolet cutoff.  No ordinary symmetry will make its
coefficient vanish.  If we imagine that the unification scale provides
the cutoff, we will find, generically, that the radiative correction
to the (mass)$^2$ is much larger than the total value we need to match
experiment.  This is an ugly situation.

In a supersymmetric theory, if the supersymmetry is not too badly
broken, it is possible to do better.  For any set of virtual particles
that might circulate in the loop there will be another graph with
their supersymmetric partners circulating.  If the partners were
accurately degenerate, the contributions would cancel.  Taking
supersymmetry breaking into account, the threatened quadratic
divergence will be cut off only at virtual momenta such that the
difference in (mass)$^2$ between the virtual particle and its
supersymmetric partner is negligible.  Notice that we will be assured
adequate cancellation if and only if supersymmetric partners are not
too far split in mass -- in the present context, if the splitting
times the square root of the fine structure constant is not much
greater than the weak scale.

\begin{figure}[!t]
\centerline{\psfig{figure=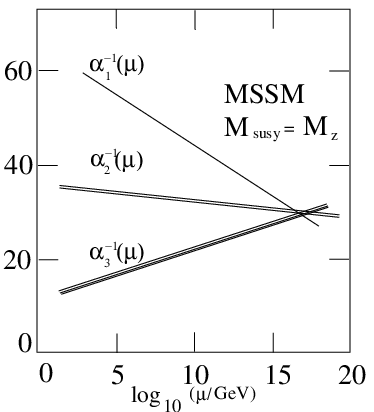,width=7cm}}
\caption[]{When the exchange of the virtual particles necessary
to implement low-energy supersymmetry, a calculation along the lines
of Figure 
\ref{fig:SM-run.eps}
comes into adequate agreement with experiment.}
\label{fig:Susy.eps}
\end{figure}

The effect of low-energy supersymmetry on the running of the couplings
was first considered long ago \cite{DRW}, in advance of the precise
measurements of low-energy couplings or of the modern limits on
nucleon decay.  One might have feared that such a huge expansion of
the theory, which essentially doubles the spectrum, would utterly
destroy the approximate success of the minimal SU(5) calculation.
This is not true, however.  To a first approximation since
supersymmetry is a space-time rather than an internal symmetry it does
not affect the group-theoretic structure of the calculation.

Thus to a first approximation the absolute rate at which the couplings
run with momentum is affected, but not the relative rates.  The main
effect is that the supersymmetric partners of the color gluons, the
gluinos, weaken the asymptotic freedom of the strong interaction.
Thus they tend to make its effective coupling decrease and approach
the others more slowly.  Thus their merger requires a longer lever
arm, and the scale at which the couplings meet increases by an order
of magnitude or so, to about 10$^{16}$ Gev.  An immediate effect of
raising the scale is to raise the mass of the gauge bosons that can
mediate proton decay, so that the experimental bounds are no longer
contradicted.  (On nucleon stability, more below.)

I want to emphasize that this very large new mass scale has emerged
unforced from the internal logic of the Standard Model itself.  It
will appear in several of our further considerations, and so for later
reference let's give it a name, the unification scale, and the token
$M_U$.

Since the running of the couplings with scale is logarithmic, the
unification of couplings calculation is not terribly sensitive to the
exact scale at which supersymmetry is broken, say between 100 Gev and
10 Tev.  It is a result robust, at the few {\it per cent\/} level,
against uncertainties of this sort.  This robustness is fortunate (and
virtually unique among the phenomenological signatures of
supersymmetry), because at present the mechanism of supersymmetry
breaking, and therefore the spectrum of sfermions and gauginos, is
quite uncertain.  The unification of couplings is also robust against
addition of additional particles, so long as they come in complete,
approximately degenerate $SU(5)$ multiplets.  Additional uncertainties
do arise from the details of the unified symmetry breaking at the high
scale.  I'll discuss these further below.  The main conclusion is that
these corrections are also at the few {\it per cent\/} level, so long
as the symmetry breaking is implemented economically ({\it i.e.},
using the simplest Higgs field representations).

On the other hand, our successful unification of couplings calculation
is most definitely {\it not\/} robust against radical changes in its
embedding framework, such as abandoning low-energy supersymmetry,
using radically different unification groups, or allowing the virtual
particles to wander off into extra dimensions.  If any of these ideas
are correct, the spectacular existing agreement of theory and
experiment, displayed in Figure 6, would seem to be a `coincidence' --
imputing to Mother Nature a rather sadistic propensity to tease.

\subsection{neutrino mass}

It is important to realize that the degrees of freedom of
the Standard Model permit neutrino masses.  A minimal implementation
of the construction requires an interaction of the type 

\begin{equation}
\Delta
{\cal L} ~=~ \kappa_{ij} L^{\alpha a i} L^{\beta b j} \epsilon_{\alpha
\beta} \phi^\dagger_a \phi^\dagger_b + {\rm h. c.} ~, 
\label{eq:newL}
\end{equation}
where $i$ and $j$ are family indices; $\kappa_{ij}$ is a symmetric
matrix of coupling constants; the $L$ fields are the left-handed
doublets of leptons, with Greek spinor indices, early Roman weak
$SU(2)$ indices, and middle Roman flavor indices; and finally $\phi$
is the Higgs doublet, with its weak $SU(2)$ index.  Two-component
notation has been used for the spinors, to emphasize that this way of
forming mass terms, although different from what we are used to for
quarks and charged leptons, is in some sense more elementary
mathematically.  $\Delta {\cal L}$ becomes a neutrino mass term when
the $\phi$ field is replaced by its vacuum expectation value $\langle
\phi^a \rangle = v \delta^a_1$.

Although this Eq. (\ref{eq:newL}) is a possible interaction for the degrees of
freedom in the Standard Model, it is usually considered to be {\it beyond\/}
the Standard Model, for a very good reason.  The new term differs from
the terms traditionally included in the Standard Model in that the
product of fields has mass dimension 5, so that the coefficient
$\kappa$ must have mass dimension -1.  In the context of quantum field
theory, it is a nonrenormalizable interaction.  When one includes it
in virtual particle loops, one will find amplitudes containing the
dimensionless factors of the type $\kappa \Lambda$, where $\Lambda$ is
an ultraviolet cutoff.  In this framework, therefore, one cannot
accept $\Delta {\cal L}$ as an elementary interaction.  It can only be
understood within a larger theoretical context.

Given a numerical value for the neutrino mass, we can infer a scale
beyond which $\Delta {\cal L}$ cannot be accurate, and degrees of
freedom beyond the Standard Model must open up.  To get oriented, let
us momentarily pretend that $\kappa$ is simply a number instead of a
matrix, and that $m = 10^{-2}$ eV is the neutrino mass.  Then, using
$v = 250$ GeV for the vacuum expectation value, we calculate 

\begin{equation}
1/M \equiv \kappa = m/v^2 = 1/(6 \times 10^{15}~ {\rm GeV}) ~~.
\label{eq:newScale}
\end{equation}

When energy and momenta of order $M$ begin to circulate in loops the
form of the interaction must be modified.  Otherwise the dangerous
factor $\kappa \Lambda$ will become larger than unity, inducing large
and uncontrolled radiative corrections to all processes, and rendering
the success of the Standard Model accidental.

Thus we trace the ``absurdly small'' value of the observed
neutrino mass scale to an ``absurdly large'' fundamental mass
scale.  You will not fail to notice that the new scale we infer here, directly 
from the observed value of the neutrino mass is, quantitatively, none other 
than $M_U$.
This is most definitely not a coincidence, as I'll now explain.  

Let us return to the question of the $N$ masses.
Because the $N^i$ are singlets, mass terms of the type
\begin{equation}
\Delta {\cal L}_N ~=~ \eta_{ij} N^{\alpha i} N^{\beta j}~ 
\epsilon_{\alpha\beta}
\label{eq:NMassTerm}
\end{equation}
with $\eta_{ij}$ a symmetric coupling matrix, are consistent with
$SU(3)\times SU(2) \times U(1)$ symmetry.  This term of course greatly
resembles the effective interaction responsible for neutrino masses,
Eq. (\ref{eq:newL}), but conceptually the difference is crucial.
Because the Ns are Standard Model singlets the Higgs doublets that
occurred in Eq. (\ref{eq:newL}) need not appear here.  A consequence
is that the operators appearing in Eq. (\ref{eq:NMassTerm}) have mass
dimension 3, so that the $\eta_{ij}$ must have mass dimension +1.
This interaction therefore does not bring in any ultraviolet
divergence problems.

What sets the scale for $\eta$?  Although Eq. (\ref{eq:NMassTerm}) is
consistent with Standard Model gauge symmetries, or even $SU(5)$, it
is not consistent with $SO(10)$.  Indeed for the product of spinor 16
we have the decomposition 16$\times$16 = 10 + 120 + 126, where only
the 126 contains an $SU(5)$ singlet component.  The most
straightforward possibility for generating a term like
Eq. (\ref{eq:NMassTerm}) in the full theory is therefore to include a
Higgs 126, and a Yukawa coupling of this to the 16s.  If the
appropriate components of the 126 acquire vacuum expectation values,
Eq. (\ref{eq:NMassTerm}) will emerge.  The 126 is a five-index
self-dual antisymmetric tensor under $SO(10)$, which may not be to
everyone's taste.  Alternatively, one can imagine that more
complicated interactions, containing products of several simpler Higgs
fields which condense, are responsible.  These need not be fundamental
interactions (they are, of course, non-renormalizable), but could
arise through loop effects or by integrating out heavier particles
even in a renormalizable field theory.

At this level there are certainly many more options than constraints,
so that without putting the discussion of N masses in a broader
context, and making some guesses, one can't be very specific or
quantitatively precise.  Nevertheless, I think it is fair to say that
these general considerations strongly suggest that $\eta$ is
associated with breaking of unified symmetries down to the Standard
Model.  Thus, if the general framework is correct, the expected scale
for its entries is set by the one we met in the unification of
couplings calculation, i.e. $\eta \sim M_U = 10^{16}$ Gev.

The Ns communicate with the familiar fermions through the Yukawa interactions 
\begin{equation}
\Delta {\cal L}_{N-L} ~=~ g^i_j {\bar N}_i L^{aj} \phi^\dagger_a~+~{\rm h.c.}~,
\label{eq:NLCoupling}
\end{equation}
using the previous notations but now, in this more conventional term,
suppressing the Dirac spinor indices.  These interactions are of
precisely the type that generate masses for the quarks and charged
leptons in the Standard Model.  If N were otherwise massless, the
effect of Eq. (\ref{eq:NLCoupling}) would be to generate neutrino
masses, of the same order as ordinary quark and lepton masses.  In
$SO(10)$, indeed, these masses would be related by simple
Clebsch-Gordon and renormalization factors of order unity.
Fortunately, as we have seen, N is far from massless.


Indeed, N is so massive that for purposes of low-energy physics we
can and should integrate it out.  This is easy to do.  The effect of
combining Eq. (\ref{eq:NMassTerm}) and Eq. (\ref{eq:NLCoupling}) and
integrating out N is to generate
\begin{equation}
\Delta {\cal L}_{\rm eff.} ~=~ g^k_i g^l_j (\eta^{-1})_{kl} L^{\alpha
a i} L^{\beta b j} \epsilon_{\alpha \beta} \phi^\dagger_a
\phi^\dagger_b + {\rm h. c.} ~.  
\label{eq:Leff}
\end{equation}
Thus we arrive back at Eq. (\ref{eq:newL}), with 
\begin{equation}
\kappa_{ij} ~=~ g^k_i g^l_j (\eta^{-1})_{kl}~. 
\label{eq:seesaw}
\end{equation}

This so-called seesaw equation \cite{seesaw} provides a much more precise 
version of
the loose connection between unification scale and neutrino mass we
discussed at the outset.  There is much uncertainty in the details,
since there is no reliable detailed theory for the $g^k_i$ nor the
$\eta$s.  But if $g$ has an eigenvalue of order unity pointing toward
the third family (this is suggested by symmetry and the value of the
top quark mass, as discussed below), and if we set the scale for
$\eta$ using the logic above, then we get close to $10^{-2}$ eV for
the $\tau$ neutrino mass, as observed.

On the face of it, then, neutrino mass of the observed magnitude
provide an additional confirmation of our well developed,
straightforward, minimalist ideas for unification beyond the Standard
Model.  It also takes us in a pretty direction where we should be
delighted to go: toward more complete symmetry, using $SO(10)$ (or
perhaps, as Pati emphasizes, a smaller but still left-right symmetric
variant).  Within this circle of ideas, neutrino mass of the observed
magnitude is a robust consequence.  Outside that circle, it becomes
another `coincidence'.

\subsection{b/$\tau$ mass ratio}

Within $SU(5)$, or any of its extensions, it is natural to expect
certain kinds of regularities among their masses, since quarks and
leptons are put together in common multiplets.  Specifically, the
right-handed $b$ quark (or, better, the left-handed $\bar b$) and the
left-handed $\tau$ lepton can be found in a single $\bar 5$ multiplet,
whereas their oppositely-handed pairs can be found in a single $10$
multiplet.  If we assume that their masses are generated in the
simplest possible way, using a Higgs field in the $\bar 5$, we find a
simple relation -- in fact, equality -- between the masses.  Such
equality does not hold, of course, of the observed physical masses.
But we must remember that -- again, in the circle of ideas around
minimalist unification -- the fundamental equality is between
effective Yukawa couplings normalized at $M_U$.  Just as for the gauge
couplings, we must renormalize this prediction down to laboratory
scales, taking into account the effect of virtual particles.  When
this is done -- again, in the minimalist framework -- one finds
striking agreement between prediction and observation.

Within $SO(10)$, one obtains (with similar assumptions) in addition
a similar
relation between the top quark mass and the underlying Dirac mass of
the $\tau$ neutrino ({\it i.e.}, the off-diagonal entry in the seesaw
mass matrix).  This reinforces the estimate of the heaviest neutrino
mass presented above, and furthermore associates that mass with the
$\tau$ neutrino.

The luster of these successes in correlating the masses of the
heaviest fermion family is somewhat dimmed by the failure of the
simplest hypotheses to explain the pattern of lighter fermion masses
and mixings.  Of course those masses, being smaller, are {\it a
priori\/} more sensitive to quantitatively small complications, so
that predictions for them are intrinsically less robust.  The
situation is far from desperate, and I'll say a bit more about it
below.

\subsection{things that don't happen}

One frequently encounters jeremiads about the danger of assuming
lack of complications such as new strongly interacting sectors
(technicolor), compositeness, and -- recently popular -- additional
large dimensions, as one extrapolates from observed energy scales to $M_U$.
Doubtless there are any number of ways that the radically conservative
extrapolation of gauge field theory might go wrong.  However, one
should not discount the observation
that similar jeremiads have been voiced for more than
twenty years now, while so far no hint of any of the suggested deviations
has in fact materialized.

Quite the contrary.  As precision measurements of Standard Model
parameters have converged on minimal supersymmetric unification of
couplings, they have also put severe constraints on these picturesque
and intuitively appealing, but speculative and phenomenologically
gratuitous, possibilities.  Likewise, searches for unconventional
sources of CP violation and for effects of neutral flavor-changing
interactions have come up empty, and put considerable pressure on any
suggestion that the fundamental dynamics associated with non-universal
flavor interactions, let alone with dynamics that connects quark and
leptons, occurs below a scale of several Tev.  Conversely, the idea
that a large scale like $M_U$ characterizes these effects, if less
tantalizing, is much safer.

There is a slightly naive, but not completely silly, general
consideration worthy of mention here.  Any ambitious extension of the
Standard Model sufficient to unify quarks and leptons (including any
incarnation of string/M theory) will almost certainly involve
violation of baryon number, and therefore at some level nucleon
instability.  With a scale as large as $M_U$, we might -- as discussed
below -- just squeeze by the experimental constraints.  If the scale
is significantly smaller, that becomes much more difficult.

\subsection{propinquity of the gravity scale}

The value of $M_U$ is, on the appropriate logarithmic scale,
remarkably close to the Planck scale $M_{\rm Planck} \sim 10^{19}~{\rm
Gev}$.  The Planck scale is the scale at which the classical Einstein
description of gravity must break down; concretely it is the energy
scale at which exchange of virtual gravitons competes quantitatively
with the other interactions.  Because $M_U$ is significantly
smaller than the Planck mass, we need not be too nervous about the
neglect of quantum gravity corrections to our unification of couplings
calculation.  Yet because it is not absurdly smaller, we can feel
encouraged for the prospect of unification including both gravity and
gauge forces, independent of any detailed model.

\subsection{broad consistency with string/M theory}

String/M theory is at present the best candidate framework for
incorporating quantum mechanics together with general relativity.
Huge challenges remain for construction of scientific world-models on
its basis.  Specifically, for example, there is no generally accepted
understanding of such basic questions as why the macroscopic world
looks 3+1 dimensional (whereas the underlying theory is more naturally
9+1 or 10+1 dimensional), nor why the cosmological term is so small,
nor even how to formulate either the basic equations or the
initial-value problem.  The rules of the game have changed over the
years, and undoubtedly will continue to do so.

Nevertheless it is intriguing that, given a sense of humor and a bit
of good will, one can descry most of the elements of reality utilized
in gauge theory unification -- the degrees of freedom of the Standard
Model, the possibility of low-energy supersymmetry, and enough
additional gauge symmetry for unification of couplings -- within
string/M theory.  The classic (vintage 1984) weakly coupled heterotic
phenomenology is mainly concerned with finding solutions to the
equations of static classical string theory that reduce well below the
Planck scale to effective theories resembling the supersymmetric
standard model.  (Of course, it is notorious that there are also
zillions of apparently equally good solutions that look nothing like
our world.)  Within this class of models, there is a large subclass
that also embodies something close to conventional gauge theory
unification.  Recent techniques support related constructions at
strong coupling \cite{hwpenn}.

In any case, I think it is certainly fair to say that there is at
present no clear contradiction between gauge theory unification as
discussed here `from the bottom up', based on straightforward
extrapolation of established facts and principles, and string/M
theory.  This tends to reinforce the significance of our previous,
model-independent result $M_U \sim M_{\rm Planck}$.

\section{Nucleon Decay}

\subsection{supersymmetry and the challenge of exotica}

I have argued for the desirability of low-energy supersymmetry based
on one major quantitative result (the unification of couplings) and
one rather soft theoretical advantage (protection of the weak scale
from radiative corrections).  Other arguments can and have been made,
but I think these two are by far the best, most concrete ones.

Against this less than overwhelming evidence we must weigh
considerable complications and several embarrassments.

In the minimal version of the Standard Model (SM), without
supersymmetry, one has the possibility of a clean, uniform explanation
of the smallness of observed CP violating effects and of neutrino
masses, and of the smallness of so far unobserved neutral flavor-changing
effects in both the quark and lepton sectors, and of nucleon
instability.  For given the symmetries and matter content of the
minimal SM, all these effects (except CP violation) arise only from
higher-dimension, nonrenormalizable interactions.  Thus they appear in
the Lagrangian multiplied by coefficients inversely proportional to
some mass scale, and if this mass scale is large (say approaching the
Planck scale) they represent unobservable, or barely observable, small
effects.

CP violation can arise through renormalizable interactions, but only
in two special ways.  One way is through complicated interference
effects involving interference among all three families, as proposed
by Kobayashi and Maskawa .  The other is through the effects of the
notorious $\theta$ term of QCD.  Existing evidence is consistent with
the idea that the first of these mechanisms is responsible for all CP
violation so far observed; while the $\theta$ term is, for a reason
presumably connected with Peccei-Quinn symmetry and the existence of
axions, very small
or zero.  The adequacy of the minimal SM framework will be tested
by future measurements of B-meson properties and searches for
elementary electric dipole moments.

As one expands the SM to include supersymmetry this clean, uniform
explanation of the absence or smallness of those many diverse species
of possible exotica comes undone.  Technically, this occurs because the
accounting of possible `relevant' (renormalizable, total mass
dimension $\leq 4$) interactions is quite different in the
supersymmetric case.  The bosonic slepton and squark fields have mass
dimension unity, as opposed to the fermionic lepton and quark fields,
which have mass dimension 3/2 (and must appear in pairs within Lorentz
invariant candidate interactions), which opens a considerably more
capacious Pandora's box.  For example, in the SM without
supersymmetry possible baryon number violating interactions have mass
dimension at least six, since to make a color singlet they must
contain at least three quark fields, and then another fermion (lepton)
to make a Lorentz singlet.  Using the squark fields, baryon-number
violating interactions with dimension 3 can be constructed.
Supersymmetry forbids these particular terms (so they may be
suppressed by the ratio of supersymmetry breaking to unification
scales), but there are several possible supersymmetric dimension 4 and
5 terms.  A complementary perspective, looking from the high scale
down, is that exchange of heavy fermion partners of scalar or gauge
fields brings in propagators with only one inverse power of the heavy
scale, instead of two, and so is less suppressed at low energy.  There
are also many additional possible sources of CP violation, no longer
necessarily involving all three families.

None of these problems appears insurmountable.  Indeed, each presents
opportunities for theoretical and experimental discovery, and each has
generated its own sizable literature.  The issue of nucleon
instability, in view of its unique sensitivity and deep cosmological
significance, may be the most critical and fundamental problem of all,
and in the remainder of this talk I will focus on it exclusively.  I
will be brief, since my collaborators will be covering some of the
same ground more thoroughly.

\subsection{from supersymmetry to Higgsino exchange}
The analysis of nucleon instability in supersymmetric theories is
difficult to discuss without introducing some technical machinery,
since supersymmetry induces some special cancellations which are
difficult to see without using superfields.  A major result of the
analysis is that the possible form of supersymmetric dimension 4
baryon number violating operators is quite restricted, and it is
easily forbidden with an appropriate discrete symmetry.  A second
major result is that the main contribution to dimension 5 baryon
number violation comes from Higgsino, not gaugino, exchange.

At first sight it might appear that the move from non-supersymmetric
unification, where the leading contributions to nucleon instability
arise from dimension 6 operators, suppressed by two inverse powers of
the unification scale, to supersymmetric unification, which allows
dimension 5 operators and nucleon instability suppressed by only one
power of the unification scale, is catastrophic.  A number of factors
mitigate this crisis, however.  The unification scale is somewhat
larger, and the relevant Higgsino mass can be larger still; the
bottom-line Higgs couplings to the light families are quite small; and
one must at the end of the day dress the scalar (squark and slepton)
fields appearing in the dimension 5 operators, by exchange of the
standard model gauginos, into ordinary quarks and leptons.

Because of all this, in order to obtain a quantitative estimate of
nucleon instability one must be quite concrete about masses of 
the superheavy
Higgs fields and their couplings to ordinary fermions.

\subsection{doublet-triplet splitting}

The Higgs doublet needed for electroweak symmetry breaking in the
Standard Model can be embedded in various ways into a representation
of the full unified gauge group.  The simplest possibility, within
$SU(5)$, is to embed it within a fundamental, {\it i.e.}, a {\bf
5}. The three extra components form a fractionally charged color
triplet.  The symmetry instructs us how this triplet couples, and we
quickly discover that it is a very dangerous object, because its
exchange violates baryon number and destabilizes nucleons.  It must be
extremely heavy, $M_{\rm triplet} \gsim 10^{14}$ Gev, in order to be
consistent with experimental limits.  In particular, it must be very
much heavier than its partner, the electroweak Higgs doublet.
Theoretically, it is quite challenging to understand how such a large
splitting could arise.  This is the doublet-triplet splitting problem.
Similar problems occur for other unification groups and embeddings.

A profound advantage of supersymmetric unification in $SO(10)$, which
in my view forms an essential adjunct to its role in protecting the
weak scale, is its ability to address the doublet-triplet splitting
problem.  Over and above its stability to radiative corrections, as
mentioned above, there is the question of obtaining the splitting at
the classical level in the first place.  There are special constraints
for the scalar potential due to supersymmetry, arising because it
comes, roughly speaking, as the square of a simpler object, the
superpotential.  They make it possible -- in $SO(10)$! -- to assure
the requisite classical splitting through a simple group-theoretic
mechanism \cite{dw}.

\subsection{fine structure of coupling unification}

By persisting in the radically conservative hypothesis that the
striking quantitative success of the unification of couplings
calculation, as displayed in Figure 6, is not accidental, we are led
to an important conclusion regarding the complexity of unified
symmetry breaking.

In general, symmetry breaking effects will split the masses of
different components of any Higgs field representation.  These
splittings lead to logarithmic changes in the running of couplings, as
mentioned above.  Let us see how they affect the fine structure of
coupling constant unification.  To lowest (one-loop) order the
modifications to the predicted value of the couplings take the form
\begin{equation}
\alpha_i^{-1} (M_Z) ~=~ \alpha_U^{-1} - {b_i\over 2 \pi} \ln (M_Z/M_U)
- \Delta_i 
\end{equation}
where 
\begin{equation}
\Delta_i ~=~ 
\sum_{{\rm submultiplets}~\kappa}  {-b_i^\kappa \over 2 \pi } \ln
(M_U/m_\kappa)~. 
\end{equation} 

Here $b_i^\kappa$ is the contribution to the $i^{\rm th}$ gauge group
$\beta$ function from the $\kappa$ submultiplet.  If all the
$m_\kappa$ are equal, the effect of the $\Delta_i$ is merely to
renormalize $M_U$.  In general, however, they will affect the
predicted relation among the observed couplngs.  Suppose that we begin
by taking all the $\Delta_i$ to vanish, which is known to lead to a
successful result.  Then if we accommodate the perturbations to
$\alpha_1^{-1}$ and $\alpha_2^{-1}$ by adjusting the two free
parameters $\alpha_U^{-1}$ and $M_U$, we are led to alter our
prediction for the strong coupling $\alpha_3^{-1}$ according to
\begin{equation}
{\delta \alpha_3  (M_Z) \over \alpha_3 (M_Z)^2} ~=~ 
{5\over 7} \Delta_1 - {12\over 7} \Delta_2 + \Delta_3~.
\end{equation}

Of course, before worrying about possible gratuitous corrections, we
must make sure that the basic fields we use to break the unified
symmetry down to the standard model give an acceptable zeroth-order
answer to begin with.  In particular, the contribution of the
electroweak doublet, and its triplet partner must be handled
carefully.  It turns out that if we let the triplet become too heavy,
say more than 100 times $M_U$, the successful zeroth-order prediction
of $\alpha_3$ becomes endangered.  Thus it is impossible to suppress
nucleon stability due to this source down to arbitrarily low levels.

Now let us estimate the quantitative impact of different kinds of
Higgs structure.  If we take a $5 + \bar 5 $ of $SU(5)$, or a 10 of
$SO(10)$, the result is
\begin{equation}
{\delta \alpha_3  (M_Z) \over \alpha_3 (M_Z)^2} ~=~ 
{1\over 2 \pi} {9\over 7} \ln (m_3/m_2).
\end{equation}
If the logarithm is of order unity, this represents a few {\it per
cent\/} correction to $\alpha_3(M_Z)$, which is tolerable.

On the other hand, consider the rank two symmetric traceless tensor 54
of $SO(10)$.  This is still one of the simpler irreducible
representations, but it contains a piece which goes as $(6,1,4/3) +
(6, 1, -4/3)$ under the standard model.  If this piece is split from
its brethren at $M_U$, the correction is
\begin{equation}
 {\delta \alpha_3  (M_Z) \over \alpha_3 (M_Z)^2} ~=~ 
{1\over 2 \pi} {51\over 7} \ln (m_3/m_U)~,
\end{equation}
which, for a logarithm of order unity, is in the neighborhood 10-20
\%.  Uncontrolled corrections of this sort could be expected to upset
the applecart.  Of course, for more complicated representations,
containing more highly charged submultiplets, the situation only gets
worse.

At face value, these considerations strongly suggest that the observed
success of the unification of couplings can be construed as reassuring
confirmation that Nature has good taste: She starts with lots of
symmetry, and uses simple, minimalistic symmetry breaking patterns.
Of course it's terribly dangerous to rely too heavily on a single
number, but we're being radically conservative, and it's taking us
where we want to go!

\subsection{a look toward fermion masses}

As we've seen, in supersymmetric unification the leading source of
nucleon instability is exchange of superheavy Higgsino fields.  In
order to pin this down, we must constrain which such fields are
present, and how they couple to quarks and leptons.  The immediately
preceding considerations strongly encourage us to restrict ourselves
to the simplest possible field content.  For $SO(10)$, concretely,
this means a small number of adjoints, fundamentals, and spinors.

Having chosen the Higgs content, we must address the coupling to
quarks and leptons.  There is, of course, a very large amount of data
regarding the masses and mixing matrices of quarks and leptons that we
should use for guidance.  Let me briefly indicate the sorts of
considerations that enter, sparing you the hairy details.  (Actually,
somewhat to my surprise, things work out rather elegantly, at least
for the second and third families.)

In the supersymmetric standard model, which we want to recover at low
energy, there are two electroweak doublets.  These emerge as the dregs
of a mass-generation process that gives superheavy masses $\sim M_U$
to all the other Higgs fields.  In general, these dregs will be made
up of bits and pieces coming from different irreducible $SO(10)$
multiplets, {\it i.e}. adjoints, fundamentals, and spinors.

Now a fundamental 10$_H$ will couple to the matter 16s by a term of
the form $g_{ij}16_i\cdot16_j\cdot10_H$, where $i,j$ are family indices and
the $g_{ij}$ are coupling constants.  Group theory requires that
$g_{ij}$ is symmetric in $i$ and $j$.  When the 10$_H$ acquires a
vacuum expectation value, these couplings will contribute to the
observable fermion mass matrices.  The group theory also correlates
the contributions to different quark and lepton mass matrices.

Similarly an effective coupling of the type
$h_{ij}16_i\cdot16_j\cdot10_H\cdot45_H$, involving an adjoint field, can
arise.  Indeed, to implement a clean gauge symmetry breaking with
doublet-triplet splitting we need a very specific form for the vacuum
expectation value of 45$_H$ (in the B-L direction).  Group theory
determines that $h_{ij}$ is antisymmetric in $i$ and $j$, and the
required alignment of the 45$_H$ introduces various factors of 3
(Georgi-Jarlskog \cite{gj} factors) into the relative contributions from this
term to quark and lepton mass matrices.

By exploiting structures of this sort, and taking guidance from
experiment, one can construct remarkably simple and overconstrained,
yet not unrealistic, models of quark and lepton masses \cite{bpw,ba,br}.

\subsection{numerical estimates; conclusion}
In our long paper, we computed the numerical consequences of a
complete model of this sort.  In constructing the model we were forced
to make several uncertain choices for the Higgs structure and
couplings, and in getting to decay rates we were forced to make
several further uncertain estimates of SUSY breaking parameters and
strong matrix elements.  I wish we could do better.  With respect to
the microscopic theory I'm afraid the situation is unlikely to improve
dramatically any time soon.  To make progress, we desperately need to
open a dialogue with Nature, through experiment.  On the other hand,
given sufficient investment in numerical QCD one could improve the
estimation of matrix elements.  That direction should certainly be
pursued, in order to insure that it will be possible to interpret
the results properly when and if they do come in.

In any case, doing the best we know how, and with all our cards on the
table, Babu, Pati and I by honest toil find
\begin{equation}
\Gamma^{-1} (p) \lsim 10^{34} {\rm yrs}.
\end{equation}  
within the circle of ideas here advocated.  The dominant modes involve
strange particles in the final state, and usually (though not
necessarily) antineutrinos.  The detailed branching ratios, and the
nature of the subdominant modes, encode information on additional
aspects of unification physics, which is very difficult to access
otherwise.

If nucleon instability at these levels were observed it would
constitute one of the greatest discoveries in the history of physics,
and provide a unique window looking out into the deep structure of
physical reality.

\end{document}